# Impact of Grid-Forming Inverters on Protective Relays: A Perspective for Current Limiting Control Design

Yifei Li, *Student Member, IEEE*, Heng Wu, *Senior Member, IEEE*, Xiongfei Wang, *Fellow, IEEE*

*Abstract*—Grid-forming (GFM) inverters can significantly alter the fault characteristics of power systems, which challenges the proper function of protective relays. This paper gives a holistic analysis of the interaction between GFM inverter-based resources (IBRs) and the supervising elements in protective relays, including directional and phase selection elements. It is revealed that the current limiting control (CLC) that is based on the current reference saturation method, adversely affects the performance of supervising elements that rely on the negative-sequence quantities. In contrast, adopting highly inductive virtual impedance in the CLC enables a reliable operation of such elements. This finding provides insights into the design of CLC for GFM IBRs from a protection perspective. It is further found that even with a highly inductive virtual impedance, the altered virtual impedance dynamics introduced by the CLC can still lead to malfunctions of the incremental quantity-based supervising elements. These theoretical findings are corroborated by simulations and controller hardware-in-the-loop (CHIL) tests.

## I. INTRODUCTION

THE supervising elements, including directional and phase selection elements, are typically implemented in protective relays for transmission line protection in synchronous generator (SG)-based power systems to determine fault directions and fault types [1], [2]. By employing supervising elements, relays are capable of tripping faulty phases in a selective manner during the forward faults, while keeping healthy phases intact. Therefore, the system reliability can be improved [3]. However, the massive integration of inverter-based resources (IBRs) are considerably changing fault characteristics of power systems, posing challenges to the reliability of supervising elements in protective relays [4].

The negative- and zero-sequence quantities are commonly used by supervising elements for protection, which, however, cannot distinguish the fault direction during a symmetrical fault, as well as differentiate the single-phase-to-ground fault from the double-phase-to-ground fault [3], [5]. To tackle this challenge, the positive-sequence quantities are often used [6], yet they are affected by the load conditions [7], [8]. To mitigate the adverse effect of load conditions, the incremental quantities are further implemented in the supervising elements [9].

The interactions between grid-following (GFL) IBRs and the supervising elements that rely on the sequence and incremental quantities are extensively studied in the literature [3] and [5]. Unlike GFL IBRs, grid-forming (GFM) IBRs operate as a slowly changing voltage source behind an impedance [10], with fault characteristics highly dependent on the current limiting control (CLC) [11], which has fundamentally different impacts on the reliability of supervising elements.

There are generally two types of CLC methods, which are the current reference saturation method and the virtual impedance method [12], [13]. The current reference saturation method can be realized through the instantaneous limiter, the priority-based limiter, or the circular limiter [11]. The instantaneous limiter cannot fully utilize the overcurrent capability of IBRs, while the priority-based limiter may keep GFM IBRs within the current-limiting mode even after the fault is cleared [14]. The circular limiter overcomes the two drawbacks, making it the most practical current reference saturation method [12].

The virtual impedance with different X/R ratios can also be directly implemented for the CLC. It is found in [15] and [16] that increasing the X/R ratio brings inherent conflicts among the small-signal stability, the transient stability, and the current limiting dynamics of GFM IBRs. In [17], it is pointed out that the GFM capability can be improved when the X/R ratio of the virtual impedance matches that of the passive impedance formed by the passive filter, the transformer, and the grid impedance, which are highly inductive in transmission grids. Therefore, using a highly inductive virtual impedance for current limitation is recommended in [18].

Extensive research works on the design of CLC for GFM IBRs are reported, but they are mostly from the perspective of stability, transient performance, and GFM capability [19], [12], [15]. Yet, the impact of CLC on the reliability of protective relays, especially supervising elements, remains an open issue. In [20] and [21], it is found that the CLC does not introduce significant impacts on distance elements. However, the findings in [20] and [21] are mainly derived from numerical simulations, which offer limited analytical insights. Moreover, the impact of CLC on the supervising elements that are based on sequence and incremental quantities has not been considered yet [22].

This paper, thus, attempts to bridge the gap by providing a systematic analysis on the impacts of different CLCs of GFM IBRs on the supervising elements employing the sequence and incremental quantities. First, based on symmetrical component theory and Kirchhoff's laws, the preconditions for the reliable operation of supervising elements are derived. The interaction between these preconditions and GFM IBRs is then examined. Consequently, analytical insights into the design of CLC for GFM IBRs from a protection perspective are provided. The main findings are summarized as follows:
1) By comparing the fault characteristics of SGs and GFM IBRs, the control dynamics of GFM IBRs that affect the reliability of supervising elements are identified. It is found that the typically used slow power control [10] has little difference from SG with respect to the impact on supervising elements. However, the CLC has a substantial impact and is significantly different from that of SG-based systems.

2) The theoretical analysis reveals that only the highly inductive virtual impedance-based CLC can guarantee the reliable operation of negative-sequence quantity-based supervising elements, while the current reference saturation methods, including the circular limiter, the instantaneous limiter, and the priority-based limiter, compromise the reliability of supervising elements and are thus unsuitable for practical applications. Moreover, it is found that with the typically used slow power control [10], the rapidly changing dynamics of IBR output impedance induced by the CLC, including both the current reference saturation and virtual impedance methods, can lead to malfunctions of supervising elements that rely on incremental quantities.

It is worth mentioning that the findings are not recognized in the prior art, as evidenced by the extensive research works on the current reference saturation methods in recent years [13], [23]. Moreover, the insights provided in this work can boost the industry's confidence in using negative-sequence quantity-based supervising elements with the highly inductive virtual impedance-based CLC. These findings also discourage reliance on incremental quantity-based supervising elements.

## II. PRECONDITIONS OF SUPERVISING ELEMENTS

This section presents the fault characteristics in single SG-based power systems, which serve as the basis for deriving the preconditions for the reliable operation of supervising elements.

### A. System Description

Fig. 1 depicts the single SG-based power system, where $i$ and $v$ represent the current and voltage at bus 1, $i_g$ and $v_e$ denote the current and voltage at bus 2, respectively. $Z_s$ and $Z_l$ correspond to the impedances of the collection transmission line and the transmission line between bus 1 and bus 2. $F_x$ and $F_y$ represent the fault points for reverse and forward faults, respectively. Distance relays $R_1$ and $R_2$ are assembled at bus 1 and bus 2. The distance relays incorporate distance elements and supervising elements, including directional and phase selection elements [1]. This work focuses on supervising elements. The *bcg* (phase *b* to phase *c* to ground) bolted fault is taken as an example here to illustrate the preconditions for supervising elements.

### B. Fault Characteristics Employed by Directional Element

Fig. 2 (a) presents the sequence network when a *bcg* fault occurs at $F_y$. $Z_g$ denotes the impedance of the grid. $v_{SG}$ and $v_g$ are the voltages of SG and the grid, respectively. The subscripts 1, 2, and 0 represent positive-, negative-, and zero-sequence quantities, respectively. The variable $m$ denotes the fault location. Applying Kirchhoff's law for Fig. 2 (a), the angle differences between voltages and currents of negative-sequence quantities ($\varphi_2$) and zero-sequence quantities ($\varphi_0$) are expressed as

$$\begin{cases} \varphi_2 = \angle \dfrac{v_2}{i_2} = \angle - \underbrace{(Z_{s2} + jX_{SG2})}_{Z_{e2}} \\ \varphi_0 = \angle \dfrac{v_0}{i_0} = \angle - \underbrace{(Z_{s0} + jX_{SG0})}_{Z_{e0}} \end{cases} \quad (1)$$

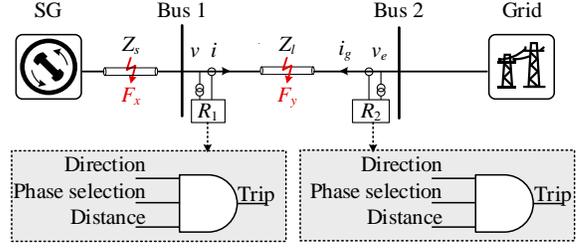

Fig. 1 Single SG-based power system.

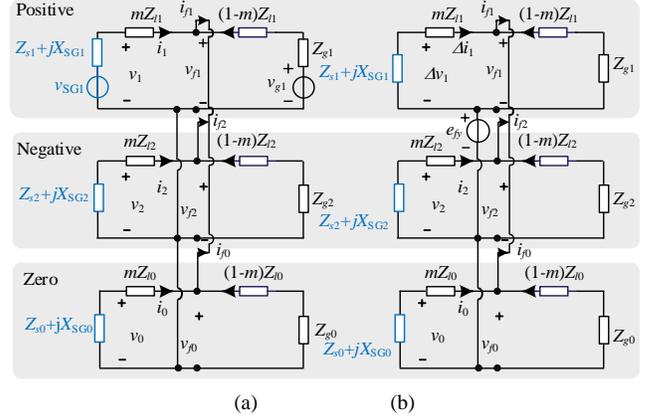

(a) (b)

Fig. 2 Sequence networks of the single SG-based power system during a *bcg* fault. (a) Sequence network. (b) Pure-fault sequence network.

where $Z_{e2}$ and $Z_{e0}$ represent the effective impedances for negative- and zero-sequence quantities, respectively. Following (1), the $\varphi_2$ and $\varphi_0$ are determined by the X/R ratios of the effective impedances, which are formed by the transmission line and SG output impedances. Since the transmission line and SG output impedances are highly inductive, both $\varphi_2$ and $\varphi_0$ approach -90° under a forward fault. In the case of a reverse fault, e.g., the fault occurs at $F_x$, $i_2$ and $i_0$ hold opposite angles to those for a forward fault. Consequently, $\varphi_2$ and $\varphi_0$ for a reverse fault are opposite in phase to those presented in (1), which approach 90°. For further details, refer to [24].

For symmetrical faults that do not involve negative- and zero-sequence quantities, the positive-sequence quantities are employed to determine the fault direction. To mitigate the influence of load conditions, the incremental, rather than total positive-sequence quantities, are employed to identify the fault direction.

Fig. 2 (b) illustrates the pure-fault sequence network for incremental quantities, where $\Delta i_1 = i_1 - i_{pre1}$, and $\Delta v_1 = v_1 - v_{pre1}$. The subscript '*pre*' stands for the pre-fault quantities. $e_{fy}$ denotes the pre-fault voltage at the fault location $F_y$. Given the negligible difference in the dynamics of $v_{SG1}$ and $X_{SG1}$ between the first few cycles after the fault inception and the time instant before fault, the SG is canceled out in Fig. 2 (b) based on superposition theory. From the positive-sequence circuit in Fig. 2 (b), the angle difference between $\Delta v_1$ and $\Delta i_1$ ($\Delta \varphi_1$) is given by

$$\Delta \varphi_1 = \angle \dfrac{\Delta v_1}{\Delta i_1} = \angle - \underbrace{(jX_{SG1} + Z_{s1})}_{Z_{e1}} \quad (2)$$

where $Z_{e1}$ is the effective impedance for positive-sequence quantities. $\Delta \varphi_1$ is also determined by the X/R ratio of the effective impedance.

Fig. 3 (a)-(c) illustrate the operation principle of directional elements that are based on incremental, negative-sequence, and zero-sequence quantities, respectively. The fault direction can be identified based on $\Delta\varphi_1$, $\varphi_2$, and $\varphi_0$. Due to the inducive nature of the SG output impedance and the transmission line impedance, the forward fault is identified when $\Delta\varphi_1$, $\varphi_2$, and $\varphi_0$ fall within the forward zone where the angle is around -90°, while the reverse fault is identified when $\Delta\varphi_1$, $\varphi_2$, and $\varphi_0$ fall within the reverse zone where the angle is around 90°. The corresponding non-operating angle is denoted as $\varphi_{non}$, which is typically set between 30° and 60° [24], and designed to improve the reliable operation of directional elements.

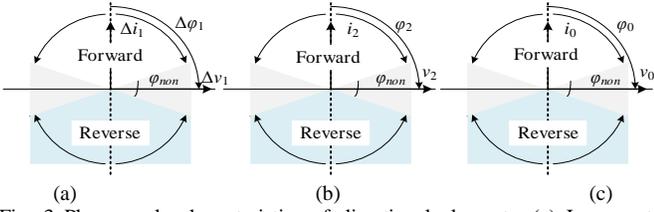

Fig. 3 Phase angle characteristics of directional elements. (a) Incremental quantity-based element. (b) Negative-sequence quantity-based element. (c) Zero-sequence quantity-based element.

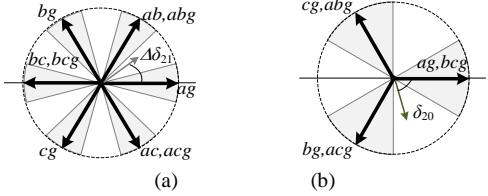

Fig. 4 The principle for PSEs. (a) $\Delta\delta_{21}$. (b) $\delta_{20}$.

### C. Fault Characteristics Employed by Phase Selection Element

The fault phases can be identified based on the one-to-one mapping between fault types and the angle differences between sequence quantities at the fault location, e.g., between $i_{f2}$ and $i_{f1}$ ($\delta_{f21}$), as well as $i_{f2}$ and $i_{f0}$ ($\delta_{f20}$) [25]. However, the relay cannot directly measure the fault-location quantities. Thanks to the highly inductive nature of the impedance in SG-based power systems, the angles of the negative- and zero-sequence quantities at the relay-assembled measurement point correspond to those at the fault location and can be expressed as

$$\begin{cases} \angle \dfrac{i_2}{i_{f2}} = \angle \dfrac{(1-m)Z_{l2}+Z_{g2}}{Z_{e2}+Z_{l2}+Z_{g2}} \\ \angle \dfrac{i_0}{i_{f0}} = \angle \dfrac{(1-m)Z_{l0}+Z_{g0}}{Z_{e0}+Z_{l0}+Z_{g0}} \end{cases} \Rightarrow \begin{cases} \angle i_2 = \angle i_{f2} \\ \angle i_0 = \angle i_{f0} \end{cases} \quad (3)$$

In contrast, the phase angles of $i_1$ and $i_{f1}$, are not necessarily equal due to the load conditions in the positive-sequence circuit. Thus, incremental quantities are used to address this mismatch issue. By applying Kirchhoff's law for the sequence network shown in Fig. 2 (b), it is derived as

$$\angle \dfrac{\Delta i_1}{i_{f1}} = \angle \dfrac{(1-m)Z_{l1}+Z_{g1}}{Z_{e1}+Z_{l1}+Z_{g1}} \Rightarrow \angle \Delta i_1 = \angle i_{f1} \quad (4)$$

Fig. 4 elaborates the principle of phase selection elements, where the angle difference between $i_2$ and $\Delta i_1$ ($\Delta\delta_{21}$), and the angle difference between $i_2$ and $i_0$ ($\delta_{20}$) are used simultaneously to identify fault types. When $\Delta\delta_{21}$ and $\delta_{20}$ fall within the fault bands, the corresponding fault types are identified. To enhance the robustness of the phase selection element, the angle bands (±15° for $\Delta\delta_{21}$, and ±30° for $\delta_{20}$), instead of fixed values, are applied to identify fault types [26].

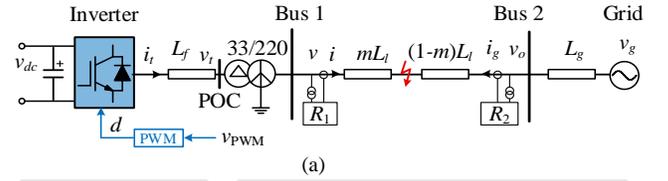
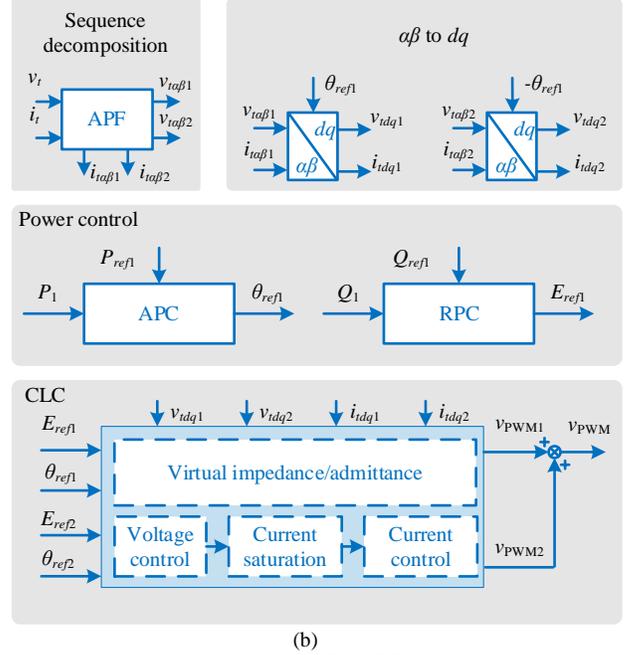

Fig. 5 Main circuit and control schemes of the GFM IBR-based power system under study. (a) Main circuit. (b) Control loops.

### D. Preconditions

Based on (1)-(4), the supervising elements operate reliably with the following preconditions:
a) The effective impedances are highly inductive.
b) The equivalent source dynamics of the SG remain nearly unchanged in the first few cycles after fault inception.

## III. IMPACTS OF GFM CONTROL ON SUPERVISING ELEMENTS

This section first presents the sequence network model of GFM IBR-based power systems, based on which the control loop that affects the reliable operations of supervising elements is identified.

### A. System Description

Fig. 5 (a) depicts the main circuit of a GFM IBR-based power system, which is used to illustrate the impacts of GFM control on supervising elements. Here, a constant DC-link voltage ($v_{dc}$) is assumed, as the DC voltage is usually taken over by a front-end converter [16]. The symbols $i_t$ and $v_t$ are the current and voltage at the point of coupling (POC), respectively. $L_f$ denotes the filter inductance. A short-circuit fault is assumed to occur at the transmission line between the bus 1 and the bus 2.

Fig. 5 (b) shows the control loops of the GFM IBR. To eliminate the second-order harmonics during asymmetrical faults, the sequence control is used, and the all-pass filter (APF) is used to decompose the positive- and negative-sequence quantities [27].

The GFM control comprises two control layers, i.e., power control and CLC. The power control employs an active power controller (APC) to synchronize the IBR with the grid, and it generates the angle reference $\theta_{ref1}$ for the *dq* transformation. Further, the reactive power control (RPC) is used to generate the voltage magnitude reference ($E_{ref1}$) for the internal voltage source. To maintain a balanced internal voltage, the reference for negative-sequence voltage magnitude ($E_{ref2}$) is set to 0 [28]. Besides the power control, the CLC methods, including current reference saturation and virtual impedance methods, are essential for IBRs to prevent overcurrent. For simplicity, the same CLC strategy is adopted for the positive- and negative-sequence quantities.

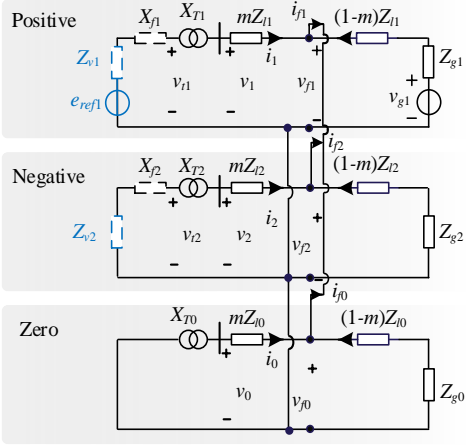

Fig. 6 Sequence network of the GFM IBR-based power system under a *bcg* fault.

### B. Sequence Network Model

Fig. 6 shows the sequence network of the GFM IBR-based power system when a *bcg* fault occurs, where the connecting manner at the fault point is determined by fault types and is not affected by the GFM control. Moreover, the Δ-Y0 transformer bypasses the direct GFM control of the IBR for zero-sequence quantities. Consequently, the zero-sequence effective impedance ($Z_{e0}$) observed from the relay-assembled point at the bus 1 only consists of the leakage impedance of the transformer, which is mainly inductive. Thus, according to (1), the supervising element relying solely on the zero-sequence quantities can operate reliably. However, such directional element is susceptible to the mutual coupling from adjacent circuits [29]. Moreover, zero-sequence quantities are absent during phase-to-phase faults.

In contrast, the positive- and negative-sequence output impedances, i.e., $Z_{v1}$ and $Z_{v2}$, are characterized by the GFM control. The exact characteristics will be detailed in Section IV. Notably, when the virtual impedance method is used and the virtual impedances are added directly to the voltage modulation reference, the filters ($X_{f1}$, $X_{f2}$) are part of the sequence network. Otherwise, they are excluded [11].

### C. Impacts of Control Loops of GFM IBRs

Fig. 7 illustrates the control schemes for APC and RPC, which determine the internal voltage source $e_{ref1}$. $K_{pR}$ and $K_{iV}$ are the proportional and integral gains for RPC, respectively. $V_{N1}$ is the nominal voltage magnitude. *D* is the P-ω droop coefficient. $K_{pP}$ and *H* represent the virtual damping and inertia constants, respectively. By employing the power control in Fig. 7, the magnitude and phase angle of the internal voltage for the positive-sequence quantities are expressed as

$$\begin{cases} E_{ref1} = \int [(Q_{ref} - Q_1)K_{pR} + V_{N1}]K_{iV} + V_1 \\ \angle e_{ref1} = \int \left[(P_{ref} - P_1)\left(\int \frac{1}{2H} + K_{pP}\right)\right] \end{cases} \quad (5)$$

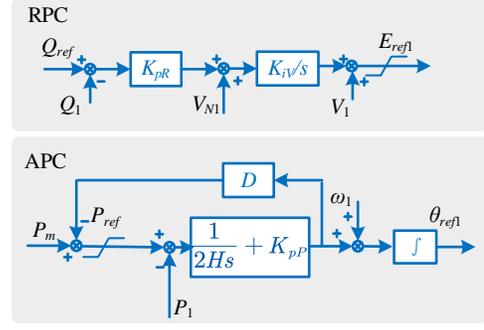

Fig. 7 Illustration of the detailed power control diagrams.

The power control adjusts $e_{ref1}$ to fulfill its function, thereby influencing the internal voltage source dynamics of the IBR. However, the output of the RPC is constrained to around 1 p.u. [10], and the bandwidth of APC is typically limited below 5 Hz through adjusting $K_{pP}$ and *H* shown in (5) [30]. Therefore, slow dynamics of the internal voltage source are anticipated for the GFM IBRs [10]. Moreover, $E_{ref2}$ is set to zero. Consequently, internal voltages for positive- and negative-sequence quantities are given by

$$\begin{cases} e_{ref1} \approx e_{pre1} \\ E_{ref2} = 0 \end{cases} \quad (6)$$

Following (6), the internal voltage sources of the GFM IBRs are like those of the SG, as shown in Fig. 2. Therefore, these two preconditions remain not significantly affected by the power control.

In contrast, the effective impedances observed from the relay-assembled point at bus 1, as defined in [18], are expressed as

$$\begin{cases} Z_{e1} = n^2 Z_{v1} + jn^2 X_{f1} + jX_{T1} \\ Z_{e2} = n^2 Z_{v2} + jn^2 X_{f2} + jX_{T2} \\ Z_{e0} = jX_{T0} \end{cases} \quad (7)$$

where *n* is the turns ratio of the transformer. When the virtual impedances are not added directly to the voltage modulation reference or the current reference saturation method is employed, the filters ($X_{f1}$, $X_{f2}$) are excluded from the effective impedances ($Z_{e1}$ and $Z_{e2}$) [11]. Based on (5), the power control primarily affects the internal voltages. With the internal voltage dynamics altered slowly, the CLC swiftly adjusts the output impedance ($Z_{v1}$ and $Z_{v2}$) to limit the fault current [12]. In this case, the effective impedance dynamics are affected based on (7). Consequently, the two preconditions are potentially influenced by the CLC.

Therefore, depending on whether the control loops affect the preconditions, the challenges for supervising elements in GFM IBR-based systems are formulated as follows:
  a) The typically used slow power control does not cause significant differences in the performance of supervising elements compared to those in SG-based power systems.
  b) The impact of CLC on the performance of both negative-sequence quantity-based and incremental quantity-based supervising elements requires further investigation.

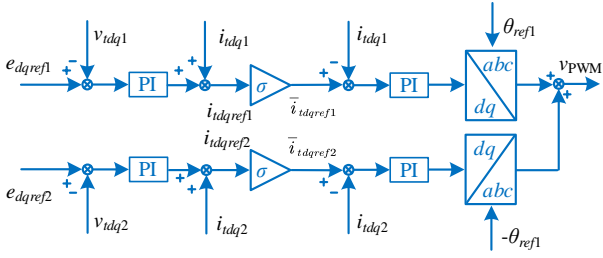

Fig. 8 The control scheme of the current saturation method.

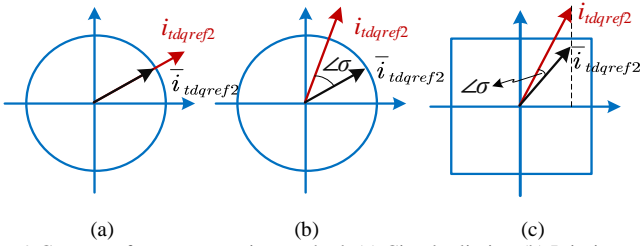

(a) (b) (c)
Fig. 9 Current reference saturation method. (a) Circular limiter (b) Priority-based limiter (c) Instantaneous limiter.

## IV. IMPACTS OF CLC METHODS ON NEGATIVE-SEQUENCE QUANTITY-BASED SUPERVISING ELEMENTS

This section examines the impacts of CLC methods on the negative-sequence quantity-based supervising elements and identifies the protection-interoperable CLC method.

### A. Current-Limiting Operation

Due to the limited current, the output impedance ($Z_{v1}$ and $Z_{v2}$) is much greater than the impedance of the power filter and the transformer, causing the X/R ratio of the output impedance to dominate that of the corresponding effective impedance. Thus, based on (1), (3), and (7), the reliability of negative-sequence quantity-based elements is ensured when the output impedance $Z_{v2}$ is mainly inductive. The X/R ratio of the output impedance of GFM IBR is thus analyzed, considering the current reference saturation and virtual impedance methods.

#### 1) Current reference saturation method

Fig. 8 shows the control scheme of the current reference saturation method [11], whose impact on the output impedance is investigated. $\sigma$ denotes the relationship between $\bar{i}_{tdqref2}$ and $i_{tdqref2}$. When the current limit is reached, the voltage integral controller is set to zero to prevent windup. Therefore, the effect of the voltage integral controller can be discarded. Moreover, the closed-loop current control is approximated as a unity gain in the analysis, considering that its bandwidth is much higher than the voltage loop, i.e., $\bar{i}_{tdqref2} = i_{tdq2}$.

Based on Fig. 8, the relationship between internal voltage and POC voltage is given by [12]

$$e_{dqref2} - v_{tdq2} = \underbrace{\frac{1-\sigma}{K_{pv}\sigma}}_{Z_{v2}} i_{tdq2} \quad (8)$$

where $K_{pv}$ is the voltage control proportional gain. Following (8), the output impedance is determined by $K_{pv}$ and $\sigma$. $\sigma$ is a complex number, which is determined by the current reference saturation method.

Fig. 9 illustrates the current reference saturation method with the circular limiter shown in Fig. 9 (a), the priority-based limiter shown in Fig. 9 (b), and the instantaneous limiter shown in

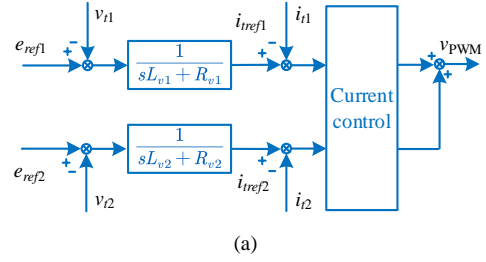

(a)

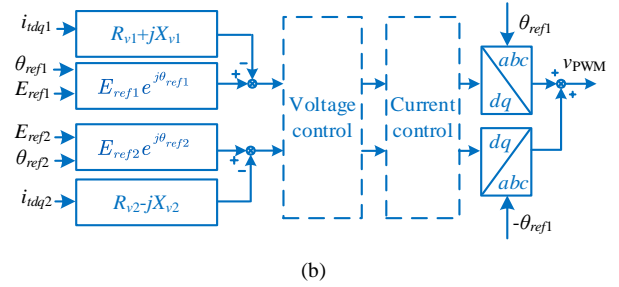

(b)
Fig. 10 Virtual impedance method. (a) Virtual admittance approach. (b) Virtual impedance approach.

Fig. 9 (c) [11]. For the circular limiter, $\angle\sigma$ is zero, and $Z_{v2}$ is mainly resistive, while $\angle\sigma$ cannot be predefined for the priority-based and instantaneous limiters. Moreover, the magnitude of $\sigma$ is affected by fault conditions and cannot be predefined either. In such cases, $Z_{v2}$ is not necessarily inductive based on (8). Consequently, the supervising elements that rely on negative-sequence quantities may malfunction with the current reference saturation method.

#### 2) Virtual impedance method

Fig. 10 presents the block diagram of the virtual impedance method, where the virtual admittance scheme is shown in Fig. 10 (a), and the virtual impedance method is shown in Fig. 10 (b), which can be implemented with or without the voltage and current control. The virtual admittance and virtual impedance for the negative-sequence quantities are, respectively, given by [10]

$$R_{v2} = \max\left(R_{vN2}, \frac{\omega_1 L_{v2}}{n_{X/R2}}\right), L_{v2} = \max\left(L_{vN2}, \frac{V_{t2}}{I_{\lim 2}\omega_1\sqrt{1+1/n_{X/R2}^2}}\right) \quad (9)$$

$$R_{v2} = \frac{X_{v2}}{n_{X/R2}}, \quad X_{v2} = \begin{cases} K_{X2}(I_{t2} - I_{th2}) & I_{t2} \geq I_{th2} \\ 0 & I_{t2} < I_{th2} \end{cases} \quad (10)$$

where $n_{X/R2}$ represents the X/R ratio. $I_{lim2}$ is the current limit. $R_{vN2}$ and $L_{vN2}$ constitute the virtual admittance in normal operation of GFM IBR. $K_{X2}$ is the proportional gain of the adaptive virtual impedance. $I_{th2}$ is the current threshold beyond which the virtual impedance is activated. Once the current limitation is triggered, the virtual impedance and the virtual admittance are defined as $Z_{v2}=R_{v2}+jX_{v2}$ and $Y_{v2}=1/(R_{v2}+j\omega_1 L_{v2})$, respectively.

Based on (1), (3), (9), and (10), implementing a small $n_{X/R2}$ adversely affects the performance of supervising elements that rely on the negative-sequence quantities. In contrast, when a sufficiently large $n_{X/R2}$ is implemented, the supervising elements can continue to operate reliably.

### B. Protection-Interoperable CLC Method

Table I summarizes the performance of supervising elements with different CLC strategies. The supervising elements that rely solely on zero-sequence quantities ($\varphi_0$) are not impacted by the CLC. However, supervising elements that rely on $\varphi_2$ and $\delta_{20}$

TABLE I SUMMARY OF SUPERVISING ELEMENTS PERFORMANCE UNDER VARIOUS CLC METHODS

| CLC | | Directional element | | | Phase selection elements | |
|---|---|---|---|---|---|---|
| | | $\Delta\varphi_1$ | $\varphi_2$ | $\varphi_0$ | $\Delta\delta_{21}$ | $\delta_{20}$ |
| Output impedance | Highly inductive | ? | ✓ | ✓ | ? | ✓ |
| | Not highly inductive | ? | ✗ | ✓ | ? | ✗ |

can only operate effectively if the effective impedance remains highly inductive.

Hence, the CLC method based on the inductive virtual impedance is identified as the protection-interoperable CLC method. All the current reference saturation methods, which introduce either an output resistor (the circular limiter), or an output impedance with an undefined impedance angle (the instantaneous limiter and the priority-based limiter), would jeopardize the reliability of supervising elements that are based on the negative-sequence quantities, and should not be employed in practice. While this conclusion is evident from a protection perspective, it remains underrecognized in literature, as evidenced by the ongoing focus on various current reference saturation methods in recent studies [22].

The CLC based on the inductive virtual impedance only guarantees reliable operations of negative-sequence quantity-based supervising elements. The impacts of CLC methods on incremental quantity-based supervising elements will be further investigated in the next section.

## V. IMPACTS OF CLC METHODS ON INCREMENTAL QUANTITY-BASED SUPERVISING ELEMENTS

This section first develops the pure-fault sequence network model for the GFM IBR-based power system. Based on this pure-fault sequence network, the impacts of CLC methods on incremental quantity-based supervising elements are analyzed.

### A. Pure-Fault Sequence Network Model

Fig. 11 presents the sequence networks under a *bcg* fault for a GFM IBR-based power system, where $Z_{eg1}=(1-m)Z_{l1}+Z_{g1}$. $Z_{eq20}$ is the equivalent parallel impedance for negative- and zero-sequence circuits. Since incremental quantity-based supervising elements rely on electrical quantities in the pure-fault sequence network for operation, deriving this sequence network is essential to assess the impact of the CLC method on these elements. Using nodal analysis, the voltage at the fault location for the fault sequence network, as shown in Fig. 11 (a), and the voltage for the pre-fault sequence network, as shown in Fig. 11 (b), are derived by

$$v_{f1} = \frac{v_1 m Y_{l1} + v_{g1} Y_{eg1}}{Y_{l1} + Y_{eg1} + Y_{eq20}} \quad (11)$$

$$e_{fy} = \frac{v_{pre1} m Y_{l1} + v_{g1} Y_{eg1}}{m Y_{l1} + Y_{eg1}} \quad (12)$$

where the symbol $Y$ stands for admittance. Subtract (11) with (12), yielding

$$\Delta v_{f1} = v_{f1} - e_{fy} = \overbrace{\frac{(v_1 - v_{pre1}) m Y_{l1}}{}}^{\Delta v_1} \frac{- e_{fy} Y_{eq20}}{m Y_{l1} + Y_{eg1} + Y_{eq20}} \quad (13)$$

Based on (13), the pure-fault sequence network for the GFM IBR-based system is obtained, as shown in Fig. 11 (c). The only difference from the SG-based power system, illustrated in

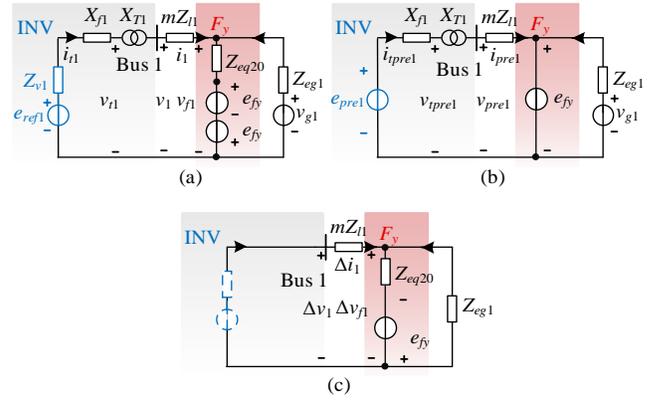

Fig. 11 Sequence networks during a *bcg* fault. (a) Fault sequence network. (b) Pre-fault sequence network. (c) Pure-fault sequence network.

Fig. 2 (b), lies in the equivalent impedance seen from Bus 1, i.e., $\Delta v_1/\Delta i_1$.

Applying Kirchhoff's law for the fault and pre-fault sequence networks, the relationship between $\Delta v_1$ and $\Delta i_1$ in Fig. 11 (c) is expressed as

$$\frac{\Delta v_1}{\Delta i_1} = \underbrace{\frac{ne_{ref1} - ne_{pre1} - n^2 i_1 Z_{v1}}{i_1 - i_{pre1}}}_{Z_{ad}} - (n^2 jX_{f1} + jX_{T1}) \quad (14)$$

In SG-based power systems, $e_{ref1} \approx e_{pre1}$, and the $I_1$ is much greater than $I_{pre1}$ [8]. From (14), it is derived $\angle Z_{ad} \approx -90°$, causing $\angle(\Delta v_1/\Delta i_1) \approx -90°$. Consequently, the circuit seen from the bus 1 can be characterized by a highly inductive impedance, as shown in Fig. 2 (b). In this case, the incremental quantity-based supervising elements can operate reliably according to (2) and (4).

However, in the GFM IBR-based system, $\angle(\Delta v_1/\Delta i_1)$ requires further investigation. Based on (14), the proximity of $\angle(\Delta v_1/\Delta i_1)$ to -90° is significantly affected by the term $Z_{ad}$. Therefore, the characteristics of $Z_{ad}$ must be thoroughly analyzed.

### B. Current-Limiting Operation

With the typical slow power control shown in Fig. 7 adopted, it is assumed that $e_{ref1} \approx e_{pre1}$. Based on (14), $Z_{ad}$ is expressed as

$$Z_{ad} = \frac{-n^2 i_1 Z_{v1}}{i_1 - i_{pre1}} \quad (15)$$

where the angle of $i_1$ does not necessarily align with that of $i_1 - i_{pre1}$. Moreover, unlike SG-based systems, $I_1$ is close to $I_{pre1}$ in GFM IBR-based systems due to the current limit. Therefore, when the current reference saturation method is employed, $\angle Z_{ad}$ cannot be predefined. Further, based on (15), even if a highly inductive $Z_{v1}$ is implemented, $\angle Z_{ad}$ may still deviate from -90°. Under these circumstances, based on (14), $\angle(\Delta v_1/\Delta i_1)$ can significantly deviate from -90°. Consequently, the incremental quantity-based supervising elements may malfunction due to the impact of CLC.

## VI. VERIFICATION RESULTS

To verify the theoretical findings, electromagnetic transient simulations and hardware-in-the-loop testing results are given in this section. The main circuit shown in Fig. 5 (a) is implemented, and its main parameters are provided in Table II. It is worth mentioning that a time delay is used with protective elements to avoid the impacts of transient variations following the fault inception [31], [32]. Moreover, control parameters can be adjusted to attenuate these transient fluctuations. Although this paper does not study the performance of supervising elements during the fluctuation period, it remains an interesting topic for future research.

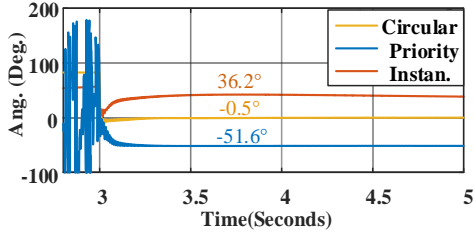

Fig. 12 Phase angles of the virtual impedance for negative-sequence quantities under current reference saturation methods.

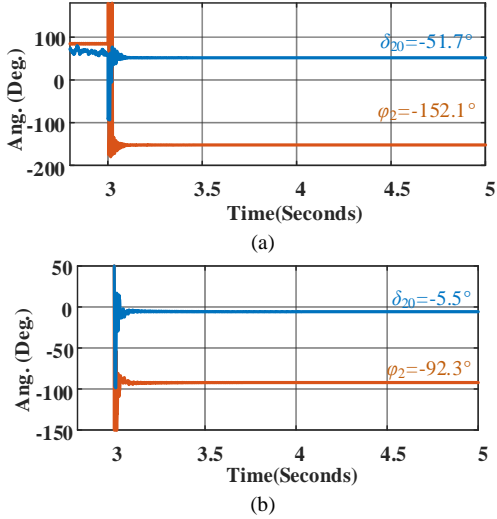

Fig. 13 Simulation results of $\varphi_2$ and $\delta_{20}$. (a) $n_{X/R2}$ =0.1. (b) $n_{X/R2}$ =20

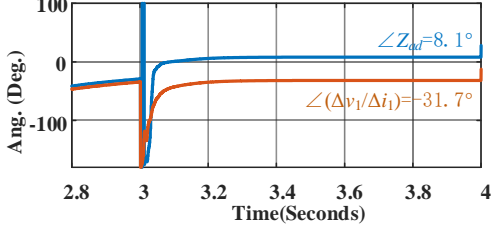

Fig. 14 Angles of $(\Delta v_1/\Delta i_1)$ and $Z_{ad}$.

### A. Simulation Results

The theatrical analysis is first validated through simulations performed on the PSCAD/EMTDC platform.

Fig. 12 illustrates the simulation results for the phase angle of the output impedance under the current reference saturation method, where the fault type considered is an *ag* fault. The phase angles of the negative-sequence output impedance under the circular, the priority-based (only the active current is injected), and the instantaneous limiters are -0.5°, -51.6°, and 36.2°, respectively. They indicate that a highly inductive output impedance is not always assured with the current reference saturation methods.

Fig. 13 shows the simulation results for $\varphi_2$ and $\delta_{20}$ with a *bcg* bolted fault occurs at $m$=0.5. In Fig. 13 (a), a low $n_{X/R2}$ of 0.1 is adopted. Under this circumstance, $\varphi_2$=-152.1° and $\delta_{20}$=-51.7° after the fault inception at 3s, both of which are out of the corresponding bands, as shown in Fig. 3 and Fig. 4. In contrast, a high $n_{X/R2}$ of 20 is adopted in Fig. 13 (b), where $\varphi_2$=-92.3° and $\delta_{20}$=-5.5°, which exactly fall into the corresponding bands.

Fig. 14 illustrates the simulation results for the case of an *ag* fault with $m$=0.01 and $R_g$=30Ω, where a highly inductive virtual impedance is triggered. $R_g$ denotes the fault resistance. Due to $\angle Z_{ad}$=4.4°, $\angle(\Delta v_1/\Delta i_1)$ deviates from -90°. Consequently, the incremental quantity-based supervising elements may malfunction, as indicated by (2) and (4).

### B. Hardware-in-the-Loop Testing Results

Fig. 15 (a) shows the controller hardware-in-the-loop (CHIL) setup that is used in this work. Three RT BOX3 units, as shown in Fig. 15 (b), are implemented with the GFM control, the IBR system model, and the relay algorithms, respectively.

Fig. 16 shows the testing results of supervising elements that are based on the negative- and zero-sequence quantities, where the correct bands for $\varphi_2$ and $\delta_{20}$ are highlighted in blue and grey, respectively. It is assumed that a bolted *bcg* fault with $m$=0.5 occurs between the bus 1 and the bus 2, with its inception time indicated by the red arrow. In Fig. 16 (a), the circular limiter is triggered. Under this circumstance, $Z_{v2}$ is resistive, which causes $\varphi_2$ and $\delta_{20}$ to fall outside the corresponding bands. In contrast, these elements operate reliably when a highly inductive virtual admittance or impedance is triggered, as shown in Fig. 16 (b) and (c), respectively.

Fig. 17 shows the testing results of the incremental quantity-based supervising elements during an *ag* fault, where $m$=0.01 and $R_g$=20Ω. It is worth mentioning that the fault resistance does not affect $\Delta\delta_{21}$ during an *ag* fault [33]. Here, the correct bands for $\Delta\varphi_1$ and $\Delta\delta_{21}$ are highlighted in blue and grey, respectively. In Fig. 17 (a), the circular limiter is triggered, making $\Delta\varphi_1$ and $\Delta\delta_{21}$ fall outside of the corresponding bands. Further, as shown in Fig. 17 (b), even with a highly inductive admittance triggered and the slow power control employed, $\Delta\varphi_1$ and $\Delta\delta_{21}$ remain outside of the corresponding bands. Fig. 17 (c) presents the results with a highly inductive virtual impedance triggered. In this case, $\Delta\delta_{21}$ deviates from the corresponding band and $\Delta\varphi_1$ approaches the boundary of the corresponding band. The results confirm the correctness of the theoretical analysis.

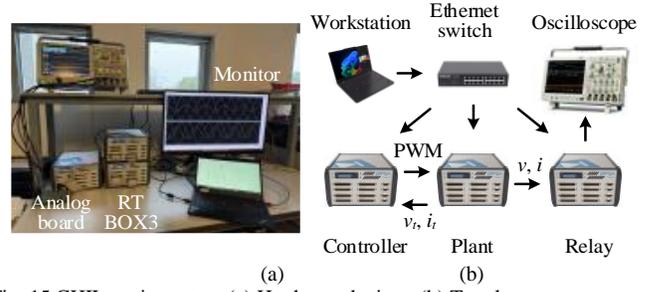

Fig. 15 CHIL testing setup. (a) Hardware devices. (b) Topology.

TABLE II PARAMETERS OF THE MAIN CIRCUITS

| Symbol | Meaning | Value |
|---|---|---|
| $v_g$ | Grid voltage (L-L, Peak) | 220kV (1.732 p.u.) |
| $f_1$ | Grid frequency | 50Hz |
| $S$ | Rated power | 100MW (1 p.u.) |
| $n$ | Turns ratio | 33kV/220kV |
| $v_{dc}$ | DC-link voltage | 40kV |
| $l$ | Length of transmission line | 100km |
| $Z_{l1}/l$ | Positive-sequence line impedance | 0.03+j0.34Ω/km |
| $Z_{l0}/l$ | Zero-sequence line impedance | 0.18+j1.19Ω/km |

### VII. CONCLUSION

In this paper, the impacts of GFM IBR on the supervising elements of protective relays, including directional and phase selection elements, are analyzed. The findings are concluded as follows:

1) The supervising elements are adversely affected by the current reference saturation method. In contrast, by using the protection-interoperable CLC method, the supervising elements that are based on the negative-sequence quantities can operate reliably.
2) Even with the adoption of typical slow power control and the protection-interoperable CLC method, the incremental quantity-based supervising elements may still malfunction. Thus, the incremental quantity-based supervising elements should not be directly used.

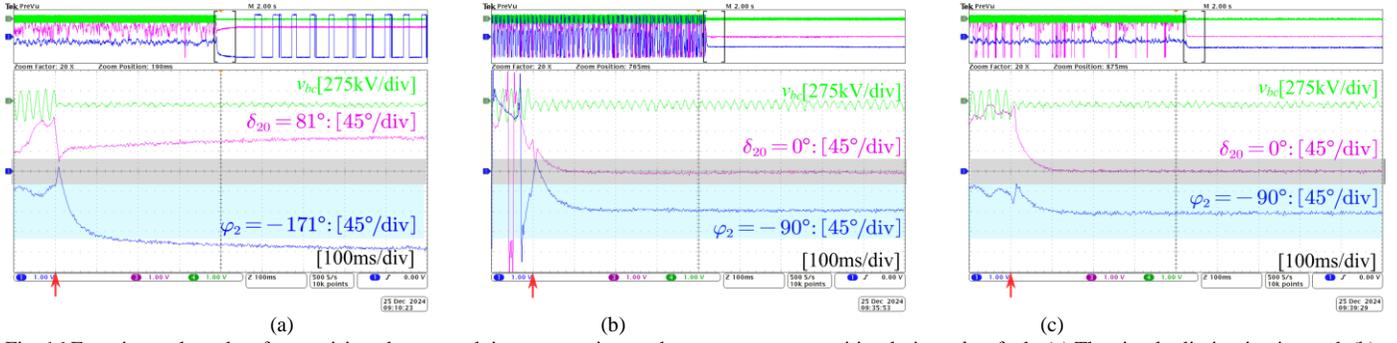

Fig. 16 Experimental results of supervising elements relying on negative- and zero-sequence quantities during a *bcg* fault. (a) The circular limiter is triggered. (b) Highly inductive virtual admittance is triggered. (c) Highly inductive virtual impedance is triggered.

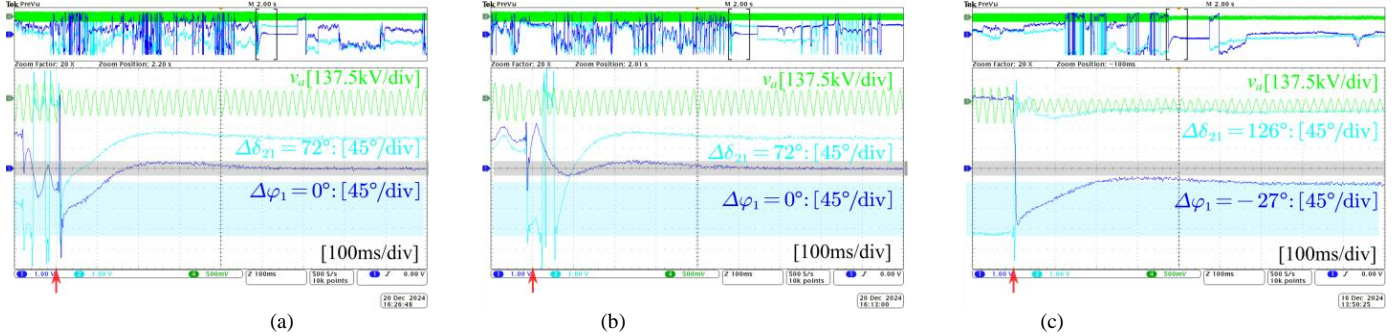

Fig. 17 Experimental results of incremental quantity-based supervising elements during an *ag* fault. (a) The circular limiter is triggered. (b) Highly inductive virtual admittance is triggered. (c) Highly inductive virtual impedance is triggered.

Simulations and CHIL tests have confirmed the findings.

## REFERENCES

[1] *D90Plus Line Distance Protection System, GE, Markham*, ON, Canada, Mar. 2012. [Online]. Available: www.gedigitalenergy.com/products/manuals/d90plus/gek-113240b.pdf
[2] R. Chowdhury and N. Fischer, "Transmission line protection for systems with inverter-based resources – Part I: Problems," *IEEE Trans. Power Del.*, vol. 36, no. 4, pp. 2416–2425, Aug. 2021.
[3] A. Hooshyar, E. F. El-Saadany, and M. Sanaye-Pasand, "Fault type classification in microgrids including photovoltaic DGs," *IEEE Trans. Smart Grid*, vol. 7, no. 5, pp. 2218–2229, Sep. 2016.
[4] B. Kasztenny, "Distance elements for line protection applications near unconventional sources," in *Proc. 75th Annu. Conf. Protective Relay Eng.*, College Station, TX, USA, Mar. 2022, pp. 1–18.
[5] Z. Yang, Z. Liu, Q. Zhang, Z. Chen, J. D. J. Chavez, and M. Popov, "A control method for converter-interfaced sources to improve operation of directional protection elements," *IEEE Trans. Power Del.*, vol. 38, no. 1, pp. 642–654, Feb. 2023.
[6] S. L. Goldsborough and A. W. Hill, "Relays and breakers for high-speed single-pole tripping and reclosing," *Electr. Eng.*, vol. 61, no. 2, pp. 77–80, Feb. 1942.
[7] S. Huang, L. Luo, and K. Cao, "A novel method of ground fault phase selection in weak-infeed side," *IEEE Trans. Power Del.*, vol. 29, no. 5, pp. 2215–2222, Oct. 2014.
[8] P. M. Anderson, *Power System Protection*. New York: IEEE Press, 1999.
[9] P. G. McLaren, G. W. Swift, Z. Zhang, E. Dirks, R. P. Jayasinghe, and I. Fernando, "A new directional element for numerical distance relays," *IEEE Trans. Power Del.*, vol. 10, no. 2, pp. 666–675, Apr. 1995.
[10] H. Wu and X. Wang, "Control of grid-forming VSCs: A perspective of adaptive fast/slow internal voltage source," *IEEE Trans. Power Electron.*, vol. 38, no. 8, pp. 10151–10169, Aug. 2023.
[11] B. Fan, T. Liu, F. Zhao, H. Wu, and X. Wang, "A review of current-limiting control of grid-forming inverters under symmetrical disturbances," *IEEE Open J. Power Electron.*, vol. 3, pp. 955–969, 2022.
[12] B. Fan and X. Wang, "Equivalent circuit model of grid-forming converters with circular current limiter for transient stability analysis," *IEEE Trans. Power Syst.*, vol. 37, no. 4, pp. 3141–3144, Jul. 2022.
[13] N. Baeckeland, D. Chatterjee, M. Lu, B. Johnson, and G.-S. Seo, "Overcurrent limiting in grid-forming inverters: a comprehensive review and discussion," *IEEE Trans. Power Electron.*, pp. 1–26, 2024.
[14] B. Fan and X. Wang, "Fault recovery analysis of grid-forming inverters with priority-based current limiters," *IEEE Trans. Power Electron.*, vol. 38, no. 6, pp. 1–10, Nov. 2022.
[15] Q. Taoufik, H. Wu, X. Wang, and I. Colak, "Variable virtual impedance-based overcurrent protection for grid-forming inverters: Small-signal, large-signal analysis and improvement," *IEEE Transactions on Smart Grid*, vol. 14, no. 5, 2023.
[16] T. Liu, X. Wang, F. Liu, K. Xin, and Y. Liu, "A current limiting method for single-loop voltage-magnitude controlled grid-forming converters during symmetrical faults," *IEEE Trans. Power Electron.*, vol. 37, no. 4, pp. 4751–4763, Apr. 2022.
[17] H. Wu, X. Wang, and L. Zhao, "Design considerations of current-limiting control for grid-forming capability enhancement of VSCs under large grid disturbances," *IEEE Trans. Power Electron.*, vol. 39, no. 10, pp. 12081-12085, Oct. 2024.
[18] ENTSOE, "Grid forming capability of power park modules," May 2024.
[19] H. Wu and X. Wang, "Small-signal modeling and controller parameters tuning of grid-forming VSCs with adaptive virtual impedance-based current limitation," *IEEE Trans. Power Electron.*, vol. 37, no. 6, pp. 7185–7199, Jun. 2022.
[20] C. L. Peralta and H. N. Villegas Pico, "Impact of Grid-Forming Converters on Distance Elements Based on the X and M Calculations," in *2024 IEEE Texas Power and Energy Conference (TPEC)*, College Station, TX, USA: IEEE, Feb. 2024, pp. 1–6.
[21] C. L. Peralta, H. P. Dang, and H. N. Villegas Pico, "Searching for Grid-Forming Technologies That Do Not Impact Protection Systems: A

[21] promising technology," *IEEE Electrific. Mag*., vol. 12, no. 2, pp. 63–70, Jun. 2024.

[22] U. Muenz, S. Bhela, N. Xue, A. Banerjee, M. J. Reno, D. Kelly, E. Farantatos, A. Haddadi, D. Ramasubramanian, and A. Banaie, "Protection of 100% Inverter-dominated Power Systems with Grid-Forming Inverters and Protection Relays – Gap Analysis and Expert Interviews", Sandia National Laboratories, SAND204-04848, 2024.

[23] T. Xu, S. Jiang, Y. Zhu, and G. Konstantinou, "Composite Power-Frequency Synchronization Loop for Enhanced Frequency Response Considering Current and Power Limits of Grid-Forming Converters," *IEEE Trans. Power Electron*., pp. 1–15, 2024.

[24] M. Benitez, J. Xavier, K. Smith, and D. Minshall, "Directional element design for protecting circuits with capacitive fault and load currents," in *Proc. 71st Annu. Conf. Protective Relay Eng*., College Station, TX: IEEE, Mar.2018, pp. 1–11.

[25] J. J. Grainger and W. D. Stevenson, *Power System Analysis. New York*: McGraw-Hill, 1994.

[26] B. Kasztenny, B. Campbell, and J. Mazereeuw, "Phase selection for single-pole tripping—Weak infeed conditions and cross-country faults," in *Proc. 27th Annu. Western Protect. Relay Conf*., Spokane, WA, USA, 2000, pp. 1–19.

[27] H. Gong, X. Wang, L. Harnefors, J.-P. Hasler, and C. Danielsson, "Admittance-dissipativity analysis and shaping of dual-sequence current control for VSCs," *IEEE Trans. Emerg. Sel. Topics Power Electron*., vol. 10, no. 1, pp. 324–335, Feb. 2022.

[28] "Specific study requirements for grid energy storage systems," FINGRID, June 2023.

[29] K. W. Jones et al., "Impact of inverter based generation on bulk power system dynamics and short-circuit performance," Task Force on Short-Circuit and System Performance Impact of Inverter Based Generation, Tech. Rep. PES-TR68, Jul. 2018.

[30] "GC0137: Minimum specification required for provision of GB grid forming (GBGF) capability (formerly Virtual Synchronous Machine/VSM Capability)" National Grid ESO, Final Modification Rep., Nov. 2021.

[31] "Technical Requirements for Interconnection to the BPA Transmission Grid," Bonneville power administration, July 2022.

[32] Z. Yang, W. Liao, Q. Zhang, C. Leth Bak, and Z. Chen, "Fault Coordination Control for Converter-interfaced Sources Compatible with Distance Protection during Asymmetrical Faults," *IEEE Trans. Ind. Electron.*, pp. 1–11, 2022.

[33] X.-N. Lin, M. Zhao, K. Alymann, and P. Liu, "Novel design of a fast phase selector using correlation analysis," IEEE Transactions on Power Delivery, vol. 20, no. 2, pp. 1283–1290, Apr. 2005.